\documentclass[pra,twocolumn,preprintnumbers,amsmath,amssymb,nofootinbib,floatfix]{revtex4}

\usepackage{graphicx,bm}
\usepackage{float}
\usepackage{subfigure}
\usepackage{amsmath}
\usepackage{amsfonts}
\usepackage{tikz}
\usetikzlibrary{shapes.geometric, calc}
\usepackage{amsthm, amssymb,bbm}
\makeatletter

\def\graphicscale{\twocolumn@sw{0.3}{0.4}}
\def\graphicthreescale{\twocolumn@sw{0.3}{0.4}}

\newcommand{\mc}[1]{\mathcal{#1}} 

\newcommand{\dt}[1]{\frac{d #1}{d t}}

\newcommand{\be}{\begin{equation}}
\newcommand{\ee}{\end{equation}}
\newcommand{\ie}{\emph{i.e.}\,}

\newcommand{\req}[1]{(\ref{#1})}
\newcommand{\tr}[1]{\mbox{Tr}[\,#1\,]}

\begin{document}
\title{On the complexity of the steady-state of weakly symmetric open quantum lattices}
\author{Davide Nigro}
\affiliation{Dipartimento di Fisica dell'Universit\`{a} di Pisa and INFN, Largo Pontecorvo 3, I-56127 Pisa, Italy}
\date{\today}

\begin{abstract}
We investigate the properties of Lindblad equations on $d$-dimensional lattices supporting a unique steady-state configuration. We consider the case of a time evolution weakly symmetric under the action of a finite group $G$, which is also a symmetry group for the lattice structure. We show that in such case the steady-state belongs to a relevant subspace, and provide an explicit algorithm for constructing an orthonormal basis of such set.
As explicitly shown for a spin-1/2 system, the dimension of such subspace is extremely smaller than the dimension of the set of square operators. As a consequence, by projecting the dynamics within such set, the steady-state configuration can be determined with a considerably reduced amount of resources.
We demonstrate the validity of our theoretical results by determinining the \emph{exact} structure of the steady-state configuration of the two dimensional XYZ model in the presence of uniform dissipation, with and without magnetic fields, up to a number of sites equal to 12. As far as we know, this is the first time one is capable of determining the steady-state structure of such model for the 12 sites cluster exactly. Altough in this work we consider explicitly only spin-1/2 systems, our approach can be exploited in the characterisation of arbitrary spin systems, fermion and boson systems (with truncated Fock space), as well as many-particle systems with degrees of freedom having different statistical properties.\\
\end{abstract}
\maketitle
\section{Introduction}
During the last decade the characterisation of open quantum systems has attracted a great deal of interest. Due to the impressive progress made in experimental physics, it is nowadays possible to control not only interactions at microscopic scales within the system under investigation, but also to design and engineer its interactions with the surrounding environment. In turn, such achievements offer the possibility of also designing the structure of steady-state configurations, that is the asymptotic configurations an open quantum system spontaneously evolves to, see \emph{e.g.} Refs. \cite{reservoirengineering0,diehl,vestraete,reservoirengineering1}. Besides fundamental aspects such as the appearence of new types of quantum transitions \cite{diss_q1,diss_q2,diss_q3,diss_q4,diss_q5,diss_q6,diss_q7,ising_trans,ising_long} or the emergence of non-trivial transport phenomena \cite{prosenznidarich1,prosenznidarich2,znidarich1,prosen1,XXZMendoza,transport_2d}, open quantum systems seem to be the prime candidates for showing the advantages of quantum simulators over classical ones \cite{preskill}. However, all these appealing perspectives come together with severe drawbacks. Indeed, predicting the dynamics of an open quantum system and determining the properties encoded into their asymptotic configurations is an hard task. This is mainly due to the the fact that contrary to the hamiltonian case, in the open case generators are not Hermitian operators. Such fact implies that the standard analytical techniques that over the years have been successfully exploited for the characterisation of closed quantum systems cannot be readily generalised to the open case.\\ 
In most of the cases the analysis of such systems is performed by means of numerical simulations. However, the numerical simulation of the dynamics occurring in open quantum system is a challenging task. Indeed, the time evolution of such systems is naturally formulated in terms of density matrices. As a consequence, in the case of open quantum systems one needs to handle objects with a typical size that grows exponentially as $\mbox{dim}[\mathbb{H}]^2$, being the  $\mbox{dim}[\mathbb{H}]$ Hilbert space dimension.\\
Such exponentially growing complexity is nowadays the primary limitation for the simulation of open systems. Nevertheless, especially for what concerns time-evolution in the Markovian limit (see Ref.\cite{petruccionebook}), it is worth noting that the efforts put by our community in the determination of new efficient algorithms for finding steady-state configurations have been payed back, and new sofisticated and promising techniques for carrying out such task have been identified (for a review on numerical methods for open quantum systems see \emph{e.g.} Ref. \cite{simulation_rev}). In the Markovian limit \cite{Lindblad,Gorini}, the time evolution is governed by master equations having the following structure ($\hbar=1$):
\be\label{eq:LGKS_equation}
\dt \rho = \mc L [\rho]= -i \left[\mc H,\,\rho\right]+\mathbb{D}[\rho],\\
\ee
where the \emph{dissipator} $\mathbb{D}[\rho]$ reads
\be
\mathbb{D}[\rho]=\sum_{j}\gamma _j \left[L_j\rho L^{\dagger}_j-\frac{1}{2}\left\{L_j^{\dagger}L_j,\,\rho\right\},\right]
\ee
being $\left[A,\,B\right]\equiv A\,B -B\,A$ and $\left\{A,\,B\right\}\equiv AB + BA$ the commutator and the anticommutator of the operators $A$ and $B$ respectively.\\
The dynamics encoded into the generator $\mc L$, usually called \emph{Louvillian} or \emph{Lindbladian}, is determined by the interplay of the unitary contributions generated by the commutator with the Hamiltonian operator $\mc H$, and non-unitary processes related to the interaction between the system and the environment encoded into the dissipator structure $\mathbb{D}[\rho]$. Such non-unitary contributions are determined by Lindblad or jump operators $\{L_j\}$, each one related to a different incoherent process controlled by the decay rate $\gamma_j$.\\

In the present work we consider master equations describing $d$-dimensional open quantum lattices on finite dimensional clusters with a \emph{unique} steady-state configuration, and \emph{weakly invariant} under the action of a finite group $G$ which is a symmetry group for the lattice structure. Our aim is to show that in such case, the system complexity is considerably reduced compared to $\mbox{dim}[\mathbb{H}]^2$. In particular, we determine the explicit structure of the subspace of square operators containing the steady-state configuration. Altough at first glance the assumptions listed here above may seem to be connected to a very specific case, it is worth noting that this is actually the usual scenario met while simulating lattice systems on finite dimensional clusters with or without boundary conditions. This is the case if one considers for instance the Ising model in transverse field \cite{ising_trans} as well as longitudinal field \cite{ising_long}, and the Bose-Hubbard model \cite{bosehubbard}, that in recent years have attracted a great deal of attention due to the perspective of modelling the non-unitary and out of equilibrium dynamics in strongly interacting many-body systems, such as Rydberg atoms systems and arrays of quantum cavities that are considered to be promising candidates for quantum simulation purposes.\\

The paper is organised as follows. In Section \ref{sec:theoretical_framework} we set the notation and discuss in detail the meaning of our assumptions. In Section \ref{sec:theoretical_results} we show a series of general theoretical results derived for the systems having the properties specified in this introductory section. In particular, by means of finite group theory, we provide an explicit algorithm for the determination of the structure of the relevant subspace containing the steady-state configuration, which is also an invariant subspace for the time-evolution. The validity of our approach is then tested in Section \ref{sec:numerical_results} by considering a case study, that is the characterisation of the magnetic properties of the steady-state configuration of the dissipative XYZ model on finite dimensional clusters, in the presence or not of external magnetic fields. In Section \ref{sec:summary_and_conclusions} we summarise our results and draw our conclusions.

\section{Theoretical framework}\label{sec:theoretical_framework}
In this work we consider a $d$-dimensional lattice system having $V$ sites. On each site we assume to have a $B$-dimensional Hilbert space $\mathbb{H}_i$, spanned by the set $\{\vert 0\rangle_i,\,\vert 1\rangle_i\,\cdots,\vert B-1\rangle_i\}$, where $i\in [1,\,V]$ is a lattice index (in what follows we denote sets of objects by means of curly brackets). In addition, we assume such set of states to be generated by the repeated action of a local operator $\hat{\alpha}^+_i$ on the reference state $\vert 0\rangle_i$, that is
\be\label{eq:single_site_number_state}
\vert n \rangle _i = a_n (\hat{\alpha}^+_i)^n \vert 0 \rangle_i,\quad n\in [0,\,B-1]
\ee
being $a_n$ a normalization factor. A lattice system locally described by the states in Eq. \ref{eq:single_site_number_state} has a total Hilbert space $\mathbb{H}$ whose dimension is $B^{V}$. Clearly, any  global lattice configuration can be expressed in terms of the states in Eq. \ref{eq:single_site_number_state}. In particular, in the following we refer to such global configurations by using the following convention
\be\label{eq:number_state}
\vert \textbf{n}\rangle = \bigotimes_{i=1}^{V}\vert n_i \rangle _i,
\ee 
where $\vert \textbf{n}\rangle$ is lattice configuration characterised by the $V$ integers contained in the entries of the vector $\textbf{n}=(n_1,\,n_2,\,\cdots\,n_V)$. In other words, in our scenario the operators $\{\hat{\alpha}^+_i\}_{i=1,V}$ and their hermitian conjugated conterparts $\{\hat{\alpha}_i\}_{i=1,V}$ play the role of creation/annihilation operators in fermions and bosons (the latter with truncated Fock space) or the role of ladder operators in spin systems. In the former case, the states in Eq. \ref{eq:number_state} represent number states, while in the latter situation they correspond to eigentates of the total magnetization along the $z$ direction.\\

As mentioned in the introductory section, we consider master equations in the Lindblad form that are $(a)$ \emph{weakly symmetric} under the action of a group $G$, which is $(b)$ a finite subgroup of the set of all the permutations of the lattice sites. In addition, we assume $(c)$ the open dynamics to be irreducible \ie the steady-state configuration is \emph{unique} \cite{evans,daviesstochastic}.\\
According to the characterisation provided in Ref.\cite{buca} (see also Ref.\cite{albert}), we say that a master equation is \emph{weakly symmetric} under the action of a unitary operator W if 
\begin{equation}\label{eq:lindbladian_symmetry}
 W\mathcal{L}[\rho] W ^{\dagger} =\mathcal{L}[W \rho W^{\dagger}].
\end{equation}
As a consequence, the assumption $(a)$ means that the condition in Eq. \ref{eq:lindbladian_symmetry} holds true for any $G_i$ belonging to the group $G$. Recall that whenever the steady-state is unique, the Lindblad equation can be only weakly symmetric. Indeed, in such case one has that the only \emph{strong symmetry} \ie the only operator commuting simultaneously with the Hamiltonian and the Lindblad operators is the identity. Notice that this is exactly the definition of irreducibility provided by Evans in \cite{evans}. The assumption $(b)$ provides some constraints on the type of transformations addressed in this paper. A finite group $G$ is a set with a finite number of elements that $(i)$ contains the identity $\mathbbm{1}$ operator, $(ii)$ for any element $G_i\in G$ also its inverse $G^{-1}_i$ belongs to $G$, and $(iii)$ for any two elements $G_l$ and $G_m$, their composition $G_l \cdot G_m $ is always an element of the group, being $``\cdot"$ the composition operation in $G$ (for an introduction to the theory of finite groups see \emph{e.g.} Refs \cite{hamermesh,rotman}). The assumption $(b)$ is equivalent to state that $G$ is a finite group where $G^{-1}_i=G^{\dagger}_i$ \ie a finite group of unitary transformations.\\

In the next section we discuss the consequences of the assumptions $(a)$, $(b)$ and $(c)$.

\section{Theoretical results}\label{sec:theoretical_results}
In this section we show a series of theoretical results, obtained for an open quantum system whose master equation satisfies the constraints listed in section \ref{sec:theoretical_framework}. In Section \ref{sec:relevant_subspace} we show the main theoretical result of this work, namely the determination of the structure of the invariant subspace containing the steady-state configuration and the description of the algorithm we used to determine an orthonormal basis of such subspace. In Section \ref{sec:parity} we discuss the role of parity, showing that whenever the generator is parity preserving, then the dynamics is fully determined by the subspace of even-parity number projectors (which is a subspace of the relevant set introduced in the previous section). Section \ref{sec:group} is devoted to a description of the symmetry group structure of a particular class of Lindbladians (notice that the dissipative XYZ model characterised in Sec. \ref{sec:numerical_results} belongs to such class).\\
\subsection{The relevant subspace and its basis}\label{sec:relevant_subspace}
Whenever the Lindblad equation is weakly symmetric under a unitary transformation $W$ and the steady-state is unique 
\be \label{eq:weak_symmetry_and_unique_ss}
W\mc L[\rho_{ss}]W^\dagger=\mc L[W\rho_{ss} W^\dagger]\Rightarrow W\rho_{ss}W^\dagger=\rho_{ss}
\ee
being $\rho_{ss}$ the density matrix associated to the steady-state configuration \ie $\mc L [\rho_{ss}]=0$. Notice that the condition reported in Eq. \ref{eq:weak_symmetry_and_unique_ss} has already appeared in literature. Indeed, in Ref. \cite{popkov} the authors exploited the presence of a weak symmetry for the determination of the  symmetry properties of the steady-state currents of a one dimensional XXZ model with dissipation at both chain ends. Nevertheless, it is worth stressing that in the present work such idea is extended to a more general scenario \ie to the case of a finite group of weak symmetries. When the Lindblad equation is weakly symmetric under a group $G$, the property reported in Eq. \ref{eq:weak_symmetry_and_unique_ss} holds true for any element of the group. As a consequence, $\rho_{ss}$ belongs to the subspace $\mc I_G$ of the operators simultaneously commuting with all the $G_i\in G$. Notice that since the unique steady-state is contained in $\mc I_G$, any configuration $\rho$ will eventually enters into this set. This is a consequence of the uniqueness of $\rho_{ss}$ (for a discussion about conditions which guaratee the attractivity of the steady-state configuration and related problems see \emph{e.g.} Refs \cite{spohn1,spohn2,frigerio1,frigerio2,spohnrev,baum1,schirmer,nigro}). Furthermore, the set $\mc I_G$ is actually an invariant subset also under the action of $\mc L$ \ie any configuration $\rho$ starting in $\mc I_G$ will remain forever inside such set. This last statement is proved in the following lines. Imagine to be able to determine an orthonormal set of matrices $\{\rho_j\}$, that is
\be\label{eq:orthonormality_of_vectors}
\tr {\rho_j^{\dagger}\rho_k}=\delta_{j\,k},
\ee
being $\delta_{j\,k}$ the Kronecker delta, which spans the entire $\mc I_G$ and whose elements do have the following property
\be\label{eq:symmetry_of_basis_set}
G_i \rho_k G_i^{\dagger}= \rho_k 
\ee  
Since $\{\rho_j\}$ is a basis for $\mc I_G$ any element of such space (even $\rho_{ss}$) has a unique decomposition in terms such set. Namely, for any $\rho \in \mc I_G$
\be\label{eq:state decomposition}
\rho=\sum_{i=j}^{\mbox{dim}[\mc I_G]}\,c_j\, \rho_j, \quad c_j\,\in \mathbb{C},
\ee
being $\mbox{dim}[\mc I_G]$ the dimension of $\mc I_G$. In order to prove that any trajectory starting at $\rho$ remains inside $\mc I_G$ it is sufficient to show that its time derivative $\mc L [\rho]$ belongs $\mc I_G$. It is easy to verify that this is the case. Indeed, due to the property in Eq. \ref{eq:symmetry_of_basis_set}, we have that 
\be
\begin{split}
G_i \mc L [\rho] G_i^{\dagger}&= \sum_{j=1}^{\mbox{dim}[\mc I_G]}\,c_jG_i\, \mc L[\rho_j] G_i^{\dagger}=\\
&=\sum_{j=1}^{\mbox{dim}[\mc I_G]}\,c_j\, \mc L[G_i\rho_j G_i^{\dagger}]=\mc L [\rho].
\end{split}
\ee
As a consequence $\mc L [\rho]$ belongs to $\mc I$. The remainder of this section is devoted to the description of an algorithm for the construction of the orthonormal set $\{\rho_j\}$.\\

Our approach is based on two steps: we first determine a partition of a basis set of square operators, then we symmetrise its elements to form the orthonormal basis mentioned before.\\
A partition of a set is a decomposition in terms of non-empty sets whose union is the initial set and their intersection is empty. In particular, we consider the following set of operators 
\be\label{eq:projectors_number_states}
\{P_{\textbf{n},\,\textbf{m}}\equiv \vert \textbf{n}\rangle  \langle \textbf{m}\vert \},
\ee
and divide them in sets where each projector belongs to one and only one set. The set in Eq. \ref{eq:projectors_number_states} is the natural basis over which operators are usually expanded when addressing lattice problems. For our purposes, we consider a partition in terms of the \emph{orbits} generated by the action of $G$ on the set in Eq. \ref{eq:projectors_number_states}, see \emph{e.g.} Ref \cite{rotman}.\\
The orbit $\mathcal{O}_{G}(P_{\textbf{n},\,\textbf{m}})$ generated by the action of the elements of $G$ on the projector $P_{\textbf{n},\,\textbf{m}}$ is the following set
\be\label{eq:orbit_definition}
\mathcal{O}_{G}(P_{\textbf{n},\,\textbf{m}})=\{X \in \mc B(\mathbb{H}):\,X=G_i P_{\textbf{n},\,\textbf{m}} G_i^{\dagger},\,G_i\in G\},
\ee
being $\mc B(\mathbb{H})$ the set of (bounded) operators acting on the Hilbert space $\mathbb{H}$. However, notice that since for any vector state $\vert \textbf{n}\rangle$ and for any $G_i\in G$ we have that 
\be\label{eq:transformation_of_number_states_group}
G_i \vert \textbf{n}\rangle  = \vert \tilde{\textbf{n}}\rangle 
\ee
being $\vert \tilde{\textbf{n}}\rangle$ another vector state, we also have that 
\be\label{eq:transformation_of_projectors}
G_i P_{\textbf{n},\,\textbf{m}}\,G_i^{\dagger}=P_{\tilde{\textbf{n}},\,\tilde{\textbf{m}}},
\ee
being $P_{\tilde{\textbf{n}},\,\tilde{\textbf{m}}}\equiv \vert \tilde{\textbf{n}}\rangle  \langle \tilde{\textbf{m}}\vert $. As a consequence, all the operators $X$ in the orbit $\mathcal{O}_{G}(P_{\textbf{n},\,\textbf{m}})$ are actually projectors having the form given in Eq. \ref{eq:projectors_number_states}.\\
At the end of this partitioning procedure we have exactly $\mbox{dim}[\mc I_G]$ different disjoint sets. Notice that, as we mention in the following section, such number can be determined \emph{a priori} (see Eq. \ref{eq:param_counting}). Now, for each orbit take one element, let us say $\bar{P}^{(j)}_{\textbf{n},\,\textbf{m}}$, being $j$ an index running on the number of orbits, and define the following operator
\be\label{eq:basis_vectors}
\rho_j \equiv \frac{1}{N_j}\sum_{s=1}^{\mbox{dim}[G]}G_s \bar{P}^{(j)}_{\textbf{n},\,\textbf{m}} G_s^{\dagger},
\ee
where $N_j$ is given by
\be
N_j=\sqrt{\tr {\rho_j^{\dagger}\rho_j}}
\ee 
Due to the fact that $G$ is a finite group, the form of the operators defined by Eq. \ref{eq:basis_vectors} is the same independently of the particular element $\bar{P}^{(j)}_{\textbf{n},\,\textbf{m}}$ chosen in the corresponding orbit. In addition, notice that the operators \req{eq:basis_vectors} are by construction $(i)$ invariant under $G$ and $(ii)$ do form an orthonormal set. This is a consequence of the fact that orbits form a partition of the orthonormal set $\{P_{\textbf{n},\,\textbf{m}}\}$.\\

\subsection{The role of parity}\label{sec:parity}
In the previous section we showed that for a system governed by a master equation with the properties listed in section \ref{sec:theoretical_framework}, the search for the steady-state configuration can be performed by analysing the dynamics within a subspace $\mc I_G$ of the entire space of operators. In this section, we show that whenever the generator $\mc L$ has an even parity, the steady-state configuration actually belongs to a subspace $\mc I^+_G$ of $\mc I_G$.\\
According to the definitions provided in section \ref{sec:theoretical_framework}, the parity operator is defined as follows
\be\label{eq:parity_operator}
\mc P \equiv \exp \left(-i \pi \sum_{s=1}^{V}\hat{\alpha}^{+}_s \hat{\alpha}_s\right)
\ee
Such operator allows us to identify two different subsets in $\mc I_G$. The first subspace, $\mc I^+_G$, which is spanned by the operators $\rho_j$ such that 
\be\label{eq:even_sector}
\mc P \rho_j \mc P^{\dagger} = \rho_j,
\ee 
contains all the operators generated by the subset formed by the projectors $P^{(j)}_{\textbf{n},\,\textbf{m}}$, for which the following quantity is an even number
\be\label{eq:delta_n_even}
\Delta N = \sum_{s=1}^{V}n_s-\sum_{t=1}^{V}m_t,
\ee
being $n_s$ and $m_s$ the integers defining respectively $\vert \textbf{n}\rangle$ and  $\vert \textbf{m}\rangle$ (see Eq. \ref{eq:number_state}). Let use denote the basis operators satisfying the constraint in Eq. \ref{eq:even_sector} by $\{\rho_j^+\}$. The complementary subspace of $\mc I^{+}_G$, denoted by $\mc I^-_G$, is generated instead by the basis operators with $\Delta N$ odd, that is $\{\rho_j^-\}$. In the same way, by considering the action of the parity operator, one can show that also the Hamiltonian operator and the Lindblad operators can be always decomposed into terms having opposite parity. In particular, we have that 
\be\label{eq:decomposition_of_operators}
\mc H = \mc H^+ + \mc H^-,\quad L_\beta =L^{+}_\beta + L^{-}_\beta,
\ee
where the even (odd) part $Q^{+(-)}$ of the operator $Q$ contains only terms with even (odd) number of operators $\{\hat{\alpha}^{+}_i\}$ and/or $\{\hat{\alpha}_i\}$ ($Q=\mc H,\,L_{\beta}$). Thanks to the decomposition in Eq. \ref{eq:decomposition_of_operators}, we have that the Lindblad equation can be decomposed as follows
\be\label{eq:Lindblad_equation_parity_decomp}
\dt \rho = \mc M^+ [\rho] + \mc M^-[\rho],
\ee
where $\mc M^+$ reads
\be\label{eq:even_parity_evolutor}
\begin{split}
\mc M^+ [\rho]&= -i \left[\mc H^+,\,\rho\right]+\sum_{\beta}\gamma_{\beta}\left[L_\beta^+\rho L_\beta^{+ \dagger}+L_\beta^-\rho L_\beta^{- \dagger}+\right.\\
&\left.-\frac{1}{2}\left\{L_\beta^{+ \dagger}L_\beta^{+}+L_\beta^{- \dagger}L_\beta^{-},\,\rho\right\} \right],
\end{split}
\ee
and being $\mc M^-$ given by
\be\label{eq:odd_parity_evolutor}
\begin{split}
\mc M^- [\rho]&= -i \left[\mc H^-,\,\rho\right]+\sum_{\beta}\gamma_{\beta}\left[L_\beta^+\rho L_\beta^{- \dagger}+L_\beta^-\rho L_\beta^{+ \dagger}+\right.\\
&\left.-\frac{1}{2}\left\{L_\beta^{+ \dagger}L_\beta^{-}+L_\beta^{- \dagger}L_\beta^{+},\,\rho\right\} \right].
\end{split}
\ee
The operator $\mc M^+$ preserves the parity of the $\{\rho_j\}$. As a consequence, it is such that 
\be\label{eq:even_parity_operator_action}
\mc M^+ [\rho^+]\in \mc I^+_G,\quad \mc M^+ [\rho^-]\in \mc I^-_G.
\ee
On the contrary, it is easy to verify that $\mc M^-$ does not preserve parity, that is
\be\label{eq:odd_parity_operator_action}
\mc M^- [\rho^+]\in \mc I^-_G,\quad \mc M^- [\rho^-]\in \mc I^+_G,
\ee
Notice that whenever $(i)$ the Hamiltonian is even and $(ii)$ the Lindblad operators have a definite parity, $\mc M^- = 0$. As a consequence, the two sectors $\mc I^+_G$ and $\mc I^-_G$ are disconnected, and correspond to invariant subspaces of $\mc I_G$. In other words, in such case the parity $\mc P$ is a weak symmetry. Therefore, the steady-state configuration belogns to $\mc I^+_G$, that is
\be\label{eq:parity_preserving_SS}
\mc P \rho_{ss} \mc P^{\dagger}=\rho_{ss},
\ee
and $\mc I^-_G$ is a decaying subspace.\\
\subsection{An example of group structure}\label{sec:group}
In this section we pay attention to the structure of the symmetry group $G$ for a particular class of open quantum lattices, namely $d$-dimensional cubic lattice systems with Hamiltonian terms and Lindblad operators that are both homogeneous and describe at most $n^{th}-$neighbors interactions. In such case, the time evolution is deteremined by a Hamiltonian operator with the following form
\be\label{eq:n_neighbor_ham}
\mc H = \frac{1}{2}\sum_{i,\,j}\left[h^{(0)}_{i,\,j}+h^{(1)}_{i,\,j}+h^{(2)}_{i,\,j}+ \cdots + h^{(n)}_{i,\,j}\right],
\ee
and by a dissipator $\mathbb{D}[\rho]$ having the following structure
\be\label{eq:sum_dissipators}
\mathbb{D}[\rho]= \sum_{k=0}^{n}\mathbb{D}^{(k)}[\rho],
\ee
with
\be\label{eq:single_dissipator}
\mathbb{D}^{(k)}[\rho]=\sum_{\beta;\,i,j} \frac{\gamma_{k,\beta}}{2}\left[L^{(k)}_{i,j;\beta} \rho L_{i,j;\beta}^{(k)\dagger}-\frac{1}{2}\left\{L_{i,j;\beta}^{(k)\dagger}L^{(k)}_{i,j;\beta},\,\rho\right\}\right],
\ee
being $h^{(k)}_{i,\,j}=h^{(k)\dagger}_{i,\,j}$ and $L^{(k)}_{i,j;\beta}$ interaction terms which couple sites $i$ and $j$ at distance $k$ ($k=0,\,1,\cdots, n$). The index $\beta$ in the Lindblad operators accounts for the possibility of having more decay channels acting on the lattice structure at distance $k$.\\

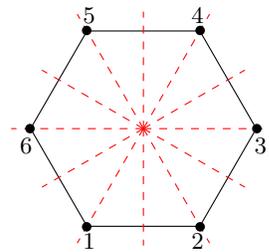
\begin{figure}
\centering
\begin{tikzpicture}[scale=3]

\def\lati{6} 
\node (hex) 
[draw, blue!0!black,rotate=0,minimum size=3cm,regular polygon, regular polygon sides=\lati] at (0,0) {};

\foreach \x in {1,2,...,\lati}{
  \draw [red, dashed, shorten <=0cm,shorten >=-0.25cm](hex.center) -- (hex.side \x);
  \draw [red, dashed, shorten <=0cm,shorten >=-0.25cm](hex.center) -- (hex.corner \x);}
  
\foreach \x in {1,2,...,\lati}
  {
  \fill (hex.corner \x) circle[radius=0.6pt];
  }

\node at (-0.52,-0.075) {6};
\node at (-0.24,+0.5) {5};
\node at (0.24,+0.5) {4};
\node at (0.52,-0.075) {3};
\node at (0.24,-0.5) {2};
\node at (-0.24,-0.5) {1};

\end{tikzpicture}
\caption{Representation of a chain with $V=6$ sites and periodic boundary conditions. Filled circles numbered from 1 to 6 correspond to lattice sites. Dashed lines correspond to the symmetry axes of the chain.}\label{fig:1d_chain}
\end{figure}

The symmetry group structure of an open quantum lattice governed by a Hamiltonian and a dissipator such those in Eq. \ref{eq:n_neighbor_ham} and Eq. \ref{eq:sum_dissipators} does depend on both the dimension $d$ and the cluster geometry, namely its shape and boundary conditions.\\
In $d=1$ and for periodic boundary conditions, since a lattice with $V=l$ sites is equivalent to a regular polygon with the same number of sides, the generator $\mc L$ is weakly invariant under the action of the dihedral group $D_V$, see \emph{e.g.} Ref.\cite{rotman}. Such group contains exaclty $2l$ elements: $l$ rotations, that correspond to lattice translations, and $l$ reflections about the $l$ different symmetry axes of the polygon. For the sake of clarity, consider a lattice chain with 6 sites. Such lattice configuration is equivalent to the hexagon in Fig. \ref{fig:1d_chain}. In the case shown in Fig. \ref{fig:1d_chain}, the geometry and the fact that interactions are homogeneous \ie they depend only on the distance between lattice sites, ensure that $\mc L$ is weakly invariant under the 6 reflection operations about the 6 different symmetry axes (red dashed lines), plus six translations that correspond to cyclic permutations of the lattice sites.
One comes to the same conclusion for any value of $V$. Indeed, after a little algebra, one finds that the Hamiltonian in Eq. \ref{eq:n_neighbor_ham} and the dissipator in Eq. \ref{eq:single_dissipator} for the $d=1$ case and periodic boundary conditions can be rewritten respectively as 
\be \label{eq:H_d1}
\mc H = \sum_{s=1}^{V} \left[ T^s \left(h^{(0)}_{1,\,1}+h^{(1)}_{1,\,2}+h^{(2)}_{1,\,3}+ \cdots + h^{(n)}_{1,\,1+n}\right) T^{s\,\dagger} \right]
\ee 
and as
\be\label{eq:dissip_1d}
\begin{split}
\mathbb{D}^{(k)}[\rho]&=\sum_{\beta}\gamma_{k,\beta}\sum_{s=1}^{V}\left[T^s L^{(k)}_{1,1+k;\beta}T^{s\,\dagger} \rho T^s L^{(k)\dagger}_{1,1+k;\beta} T^{s\,\dagger}\right.\\
&\left. -\frac{1}{2}\left\{T^s L^{(k)\dagger}_{1,1+k;\beta}L^{(k)}_{1,1+k;\beta}T^{s\,\dagger},\,\rho\right\}\right],
\end{split}
\ee
being $T$ the generator of translations in $d=1$, satisfying $T^{V}=\mathbbm{1}$. For what concerns the weak invariance under translations, it is sufficient to observe that translations of the lattice structure belong to the finite subgroup of rotations of $D_V$. Therefore, since in Eq.\ref{eq:H_d1} and Eq. \ref{eq:dissip_1d} appear all the $V$ elements of such subgroup, the action of $T$ simply permutes the order of the terms in Eq.\ref{eq:H_d1} and Eq. \ref{eq:dissip_1d}. More explictly, one finds that 
\be
T \mc H T^{\dagger}= \mc H,\quad T \mathbb{D}^{(k)}[\rho] T^{\dagger}=\mathbb{D}^{(k)}[T \rho T^{\dagger}],
\ee
which clearly implies the weak invariance of the open dynamics. For what concerns instead the symmetry under reflections, it is sufficient to note that such transformations map first neighbor sites into first neighbor sites. For the same reasons discussed in the lines above, such fact ensures that the dynamics is also weakly invariant under reflections. Further, notice that all the elements in $D_V$ can be written as a the product of a translation and of a reflection, see \emph{e.g.} Ref. \cite{rotman}.\\

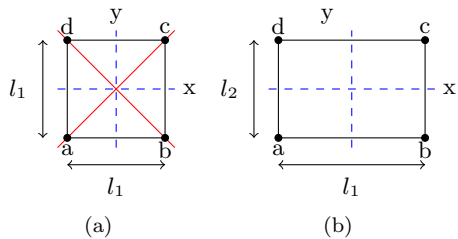
\begin{figure}[htb]
\centering
\subfigure[]{
\begin{tikzpicture}[scale=0.65]
\draw (0,0)--(2,0)--(2,2)--(0,2)--(0,0);

\draw [red] (-0.2,-0.2)--(2.2,2.2);
\draw [red] (-0.2,2.2)--(2.2,-0.2);

\draw [blue,dashed] (-0.2,1)--(2.2,1);
\node at (2.5,1) {x};
\draw [blue,dashed] (1,-0.2)--(1,2.2);
\node at (1,2.5) {y};

\filldraw (0,0) circle (2pt);
\filldraw (2,0) circle (2pt);
\filldraw (2,2) circle (2pt);
\filldraw (0,2) circle (2pt);

\node at (0,-0.25) {a};
\node at (2,-0.25) {b};
\node at (2,2.25) {c};
\node at (0,2.3) {d};

\draw [<->] (0,-0.55)--(2,-0.55);
\draw [<->] (-0.5,0)--(-0.5,2);
\node at (-1,1) {$l_1$};
\node at (1,-1) {$l_1$};
\end{tikzpicture}\label{fig:square}}
\subfigure[]{\begin{tikzpicture}[scale=0.65]
\draw (0,0)--(3,0)--(3,2)--(0,2)--(0,0);

\draw [blue,dashed] (-0.2,1)--(3.2,1);
\node at (3.5,1) {x};
\draw [blue,dashed] (1.5,-0.2)--(1.5,2.2);
\node at (1,2.5) {y};

\filldraw (0,0) circle (2pt);
\filldraw (3,0) circle (2pt);
\filldraw (3,2) circle (2pt);
\filldraw (0,2) circle (2pt);

\node at (0,-0.25) {a};
\node at (3,-0.25) {b};
\node at (3,2.25) {c};
\node at (0,2.3) {d};

\draw [<->] (0,-0.55)--(3,-0.55);
\draw [<->] (-0.5,0)--(-0.5,2);
\node at (-1,1) {$l_2$};
\node at (1.5,-1) {$l_1$};
\end{tikzpicture}\label{fig:rectangle}}
\caption{Sketch of a $l_1 \times l_2$ cluster. Fig. \ref{fig:square}: representation of a square cluster. The four circles labelled by a, b, c and d correspond to the four sites in the corners of the cluster. Solid and dashed lines correspond to the symmetry axes of the square cluster. Fig. \ref{fig:rectangle}: representation of a rectangular cluster. The four circles labelled by a, b, c and d correspond to the four corners of the cluster. Dashed lines correspond to the symmetry axes of the rectangular cluster.}\label{fig:corner_plus_symmetry}
\end{figure}

In $d \geq 2$, the symmetry group structure is more complex, so in what follows we focus on the $d=2$ case. Analogous results can be found in higher dimensions. As in the $d=1$ case, in $d=2$ the symmetry group $G$ of a $l_1 \times l_2$ cluster is defined by the elements of the subgroup of translations $\mc T$, and by those belonging to the subgroup $\mc S$ that contains information about reflectional and rotational symmetries of the lattice cluster. In particular, as we show in the following, any element of $G$ can be written as the composition of a translation with a reflection or a rotation. For what concerns the subgroup $\mc T$, we have that it is generated by two operators $T_1$ and $T_2$. The presence of periodic boundary conditions imposes the following constraint 
\be\label{eq:PBC_constraint}
T_k^{l_k}=\mathbbm{1}\quad (k=1,2)
\ee 
In particular, any element of $\mc T$ can be uniquely written as a product of powers of the group generators. Namely, we have that 
\be\label{eq:translations_decomposition}
\mc T = \{T_{\alpha \beta}:\,T_{\alpha \beta}\equiv T_1^{\alpha}T_2^{\beta},\quad \alpha\in[
1,l_1],\beta\in[1,l_2]\}
\ee
For what concerns the structure of $\mc S$, we have two possibilities, depending on whether or not $l_1=l_2$. To this purpose, please pay attention to Fig. \ref{fig:corner_plus_symmetry}.
If $l_1=l_2$ (see Fig. \ref{fig:square}) the subrgroup $\mc S$ coincides with the symmetry group of the square, that is $D_4$. Such group contains the following eight elements
\be\label{eq:S_elements_square}
\mc S=\{\mathbbm{1},\,D_{ac},\,D_{bd},\,R_x,\,R_y,\,\Omega,\,\Omega^2,\,\Omega^3\}
\ee
that correspond respectively to the identity, a reflection with respect to the a-c diagonal, a reflection with respect to the b-d diagonal, a reflection with respect to the $x$ axis, a reflection with respect to the $y$ axis, and a $\pi /2$, $\pi$, $3\pi /2$ counterclockwise rotation about the center of the cluster.\\ 
If $l_1\neq l_2$ (see Fig. \ref{fig:rectangle}), the subgroup $\mc S$ describes instead the symmetry group of the rectangle, that is 
\be\label{eq:S_elements}
\mc S=\{\mathbbm{1},\,R_x,\,R_y,\,\Omega^2\}
\ee
As mentioned above, any element of the symmetry group $G$ can be written as the composition of a translation with an element belonging to $\mc S$. In order to see this, one needs to show that for any $S_{\gamma}\in \mc S$ and $T_{\alpha\beta}\in \mc T$ there exist two operators $S_{\bar{\gamma}}$ and $T_{\bar{\alpha}\bar{\beta}}$ such that 
\be\label{eq:reordered_products_thesis}
S_{\gamma}T_{\alpha\beta}=T_{\bar{\alpha}\bar{\beta}} S_{\bar{\gamma}}
\ee
To prove the validity of Eq. \ref{eq:reordered_products_thesis}, it is sufficient to consider the action of the elements of $\mc S$ on those belonging to $\mc T$, that is to consider for any $T_{\alpha \beta} \in \mc T$ the following operator
\be\label{eq:T_tilde_transf}
O_{\gamma,\,\alpha,\,\beta}=S_{\gamma}  T_{\alpha\beta} S_{\gamma}^{\dagger},\quad S_{\gamma}\in \mc S.
\ee
Independently of $l_1$ and $l_2$, the group structure ensures that $O_{\gamma,\,\alpha,\,\beta}\in G$. Most importantly, since reflections and rotations map translations into other translations, one obtains that $O_{\gamma,\,\alpha,\,\beta}\in \mc T$. As a consequence, as stated above, we have that
\be\label{eq:reordered_products}
S_{\gamma}T_{\alpha\beta}=O_{\gamma,\,\alpha,\,\beta} S_{\gamma}
\ee
The weak invariance of the Lindbladian under the action of the group $G$ in the present case, as well as in higher dimensions, can be proved by following the same procedure discussed before for the $d=1$ case.\\

According to the theoretical description shown in the previous sections, provided that the steady-state configuration is unique, it is contained inside the manifold $\mc I _{G}$. Clearly, since the configurations spanning $\mc I_G$ are constructed by considering the orbits generated by $G$, whose structure is determined by the dimension $d$ and by the geometry, also the dimension of such subspace varies with such parameters. Nevertheless, the dimension of such subspace can be computed \emph{a priori}. Indeed, by means of the Burnside's Lemma (see also  Poly\'{a}'s Theorem) \cite{rotman}, one finds that the dimension of $\mc I_{G}$ is given by
\be\label{eq:param_counting}
\mbox{dim}[\mc I _{G}]= \frac{1}{\vert G \vert}\sum_{G_i\in G}\vert \mc P ^{G_i}\vert, 
\ee
where for a given set $A$, $\vert A \vert$ denotes its cardinality (that is the number of its element), and being $ \mc P^{G_i}$ the \emph{stabilizer set} of $G_i$, that is
\be\label{eq:stabilizer}
\mc P^{G_i}=\{P_{\textbf{n},\,\textbf{m}}:\,  G_i\,P_{\textbf{n},\,\textbf{m}}G^{\dagger}_i=P_{\textbf{n},\,\textbf{m}}\}
\ee
In order to exemplify the advantages of our approach, we report in Table \ref{tab:parameter_counting} some results for a two dimensional system of spin-1/2 particles on $l_1\times l_2$ clusters.
\begin{table}
\begin{tabular}{|c|c|c|c|}
\hline
$\,l_1 \times l_2$ \,&\,  $\mbox{dim}[\mathbb{H}]^2$\,&\, $\mbox{dim}[\mc I_G]$\, &\, $\mbox{dim}[\mc I^+_G]$\, \\\hline
$2 \times 2$ &$2^8=256$         & 55 	& 31 		\\
$2 \times 3$ &$2^{12}=4096$     & 430   & 226 		\\
$2 \times 4$ &$2^{16}=65536$    & 4756  & 2452 		\\
$3 \times 3$ &$2^{18}=262144$   & 4480  & 2240 		\\
$2 \times 5$ &$2^{20}=1048576$  & 53764  & 27036 	\\
$3 \times 4$ &$2^{24}=16777216$ & 367645  & 184341 	\\
\hline
\end{tabular}
\caption{Number of parameters needed for determining the steady-state configuration of a $l_1 \times l_2$ cluster of a spin-1/2 system. The parameter $\mbox{dim}[\mathbb{H}]^2=2^{2 V}$ provides the dimension of a square matrix for a two-level system on a $l_1 \times l_2$ cluster \ie $2^{2(l_1 \times l_2)}$. The parameter $\mbox{dim}[\mc I_G]$ represents the maximun number of parameters required in absence of other symmetries. If the generator $\mc L$ is parity-preserving, then the steady-state is contained in a subspace of dimension $\mbox{dim}[\mc I^+_G]$.}\label{tab:parameter_counting}
\end{table} 
There we report for a configuration of size $l_1 \times l_2$ (first column), $\mbox{dim}[\mathbb{H}]^2$ (second column) which provides the order of magnitude for the total number of parameters usually required for representing such type of problems, the dimension of $\mc I_G$ (third column), and finally the dimension of $\mc I^+_G$ (fourth column). As one can see, the number of parameters $\mbox{dim}[\mc I_G]$ needed for the determination of the steady-state configuration is considerably smaller than the total number of parameters needed in the representation of a square operator. In addition, recall that $\mbox{dim}[\mc I_G]$ is the \emph{maximum} number of parameters required. Therefore, in the presence of other symmetries (compatible with the lattice symmetries), the steady-state configuration actually belongs to a subspace with dimension smaller that $\mbox{dim}[\mc I_G]$. This is the case, if the Lindbladian is parity-preserving. In such case, as we show in Sec. \ref{sec:parity_broken}, the steady-state actually belongs to $\mc I^+_G \subset \mc I_G$. \\
\section{Numerical results}\label{sec:numerical_results}
In this section we show some results that corroborate the theoretical framework discussed in the previous sections. In Section \ref{sec:chi}, we investigate the magnetic properties of the dissipative XYZ model on finite dimensional clusters. Here, we show that by means of the algorithm discussed in the previous sections we obtain results in agreement with those previously appeared in literature. In addition, we show \emph{exact} numerical results for such model up to $V=12$. As far as we know, this is the first time that exact results for the steady-state configuration of the open XYZ model on a 12 sites cluster have been derived. In Section \ref{sec:parity_broken} we pay attention to the role of parity. To this puprose, we discuss the effects on the steady-state structure of a local uniform magnetic field in the $xy$ plane. Such interaction term mixes sectors with opposite parity. Therefore, a different structure of the steady-state configuration should be detectable.
\subsection{Magnetic properties of the XYZ model}\label{sec:chi}
Until now we discussed the role of symmetries in general terms. Here, we focus on a particular case study, namely the open XYZ model on $d$-dimensional cubic clusters with periodic boundary conditions and uniform dissipation. The generator we considered reads ($\hbar=1$)
\be\label{eq:dissipative_XYZ}
\begin{split}
\mc L_0[\rho]=&-i\sum_{\langle i,j\rangle}\left[J_{x}\sigma_{i}^{x}\sigma_{j}^{x}+J_{y}\sigma_{i}^{y}\sigma_{j}^{y}+J_{z}\sigma_{i}^{z}\sigma_{j}^{z},\rho\right] +\\
&+\gamma\sum_{i=1}^{V}\left[\sigma_{i}^-\rho\sigma_{i}^+-\frac{1}{2}\left\{\sigma_{i}^+\sigma_{i}^-,\,\rho\right\}\right]
\end{split}
\ee
The term in the first line of Eq. \ref{eq:dissipative_XYZ} accounts for unitary processes related to the commutator with the XYZ Hamiltonian. Such operator describes the interactions between spin-1/2 particles located at first-neighbor sites ($\langle i ,\,j\rangle$ denotes first-neighbor sites). The strength of such interaction is controlled by the three couplings denoted by $\{J_{\alpha}\}$ ($\alpha=x,\,y,\,z$). As usual, such particles are described by means of a set of Pauli operators and the corresponding ladder operators, denoted respectively by $\sigma_{i}^{\alpha}$ ($\alpha=x,y,z$) and by $\sigma_{i}^{+}$ and $\sigma_{i}^{-}$. 
The terms in the second line of Eq. \ref{eq:dissipative_XYZ} account for the non-unitary processes. In particular, in our framework each spin is locally and incoherently driven by a bath at a rate controlled by the parameter $\gamma$. Notice that the structure of the generator $\mc L_0$ is exactly as the one considered in Section \ref{sec:group}, with $\mc H$ having only first neighbor interactions and the dissipator having $V$ local decay channels.\\

Independently of the dimension $d$, of the system volume $V$ and of the couplings $\{J_{\alpha}\}$, the generator (\ref{eq:dissipative_XYZ}) has a unique fixed point $\rho_{ss}$ \ie the steady-state is unique \cite{nigro}. As any fixed point, such configuration in contained into the nullspace or Kernel of the generator $\mc L_0$. Therefore, one can determine its structure by means of linear algebra techniques, such those exploited in the ARPACK library \cite{arpack}, a linear algebra library suitable for finding few eigenvalues and the corresponding eigenvectors of large and sparse matrices.\\

As a proof of the validity of our approach, we investigate the magnetic properties of the dissipative XYZ model. Recently, such problem has been addressed by means of different techniques (see \emph{e.g.} Refs. \cite{lee,jin2,rota,tensornet,rotamc,nagy,casteels,net1,net2}). However, in what follows we mainly refer to the results shown in Ref. \cite{rota}, where the susceptibility properties of the dissipative XYZ model related to the application of a magnetic field in the $xy$ plane have been investigated in $d=2$.\\
The Louvillian in Eq. \ref{eq:dissipative_XYZ}, and as a consequence its steady-state, has a $\mathbbm{Z}_{2}$ symmetry in the $xy$-plane which implies that the following expectation values
\be\label{eq:sx_sy_expectations}
\langle \sigma^{\alpha}_{i}\rangle = \tr { \sigma ^{\alpha} _{i} \rho _{ss}},\,\quad (\alpha = x,\,y)  
\ee
and in general expectation values involving an odd number of $\sigma^{x}_{i}$ and/or $\sigma^{y}_{i}$ are identically zero.\\
In the presence of a non-zero magnetic field $\vec{h}$ in the $xy$-plane such symmetry is explicitly broken, giving rise to non-zero expectation for the spin magnetisation in the $x$ and $y$ directions. The presence of such field can be modelled by adding the following term to XYZ hamiltonian 
\begin{equation}\label{eq:magnetic_field}
\mc H _{m} = \sum_{i=1}^{V}\left[h_x\sigma^{x}_{i}+h_y\sigma^{y}_{i}\right],
\end{equation}
where $\{h_{\alpha}\}$ ($\alpha=x,y$) denote the components of the magnetic field.\\
When the intensity of the magnetic field is sufficiently small ($\vert \vert \vec{h}\vert \vert = \sqrt{h_x^2+h_y^2}\ll 1$), we expect to have a local magnetization $\vec{M}$ linearly dependent on the components of the external field. In other words, we expect to observe the following behavior
\begin{equation}
M_{\alpha}=\sum_{\beta = x,\,y}\chi_{\alpha \beta}h_{\beta} + o(\vert \vert \vec{h} \vert\vert^2 ),\quad (\alpha=x,\,y)
\end{equation}
being $M_{\alpha}\equiv\langle \sigma^{\alpha}_{i}\rangle$, and being $\chi_{\alpha \beta}$ the $\alpha-\beta$ component of the suceptibility tensor, whose explicit expression reads
\begin{equation}\label{eq:chi_components}
\chi_{\alpha,\,\beta}=\left.\frac{\partial M_{\alpha}}{\partial h_{\beta}}\right\vert_{ \vert \vert \vec{h} \vert\vert^2 = 0}
\end{equation}
Notice that due the homogeneity of the magnetic field, the response is uniform \ie $\vec{M}$ does not depend on the lattice site.\\
\begin{figure}
\centering
\includegraphics[scale=0.6]{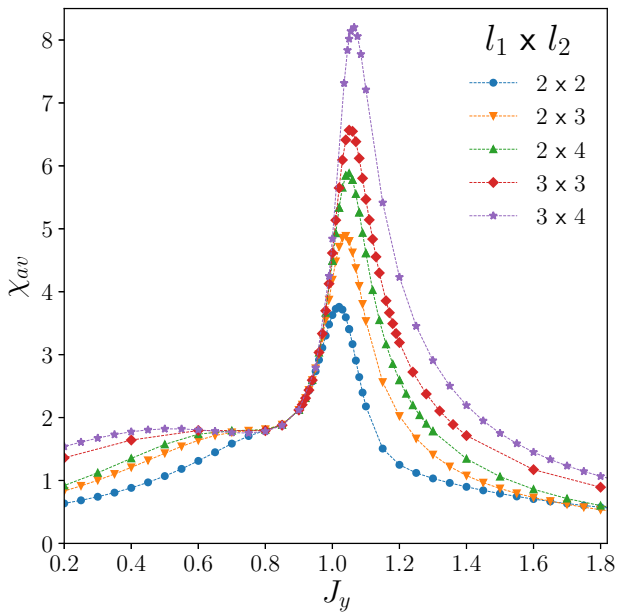}
\caption{Behavior of the angularly averaged susceptibility $\chi_{av}$ at increasing cluster volume for $J_x=0.9$, $J_z=1$, $\gamma=1$ and different values of $J_y$. The cluster size $l_1 \times l_2$ is visible in the legend. All the quatities are expressed in units of $\gamma$.}\label{fig:chi_average}
\end{figure}
The susceptibility tensor has been exploited in Ref. \cite{rota} to define another quantity of interest, that is the \emph{angularly averaged susceptibility} whose expression reads
\begin{equation}\label{eq:angularly_averaged_X}
\chi_{av}=\frac{1}{2\pi}\int_{0}^{2\pi}d\theta \left.\frac{\partial\vert \vec{M}(h,\,\theta)\vert}{\partial h}\right\vert_{\vert\vert\vec{h}\vert\vert^{2}=0},
\end{equation}
being 
\begin{equation}\label{eq:abs_of_M}
\left.\frac{\partial\vert \vec{M}(h,\,\theta)\vert}{\partial h}\right\vert_{\vert\vert\vec{h}\vert\vert^{2}=0}=\left\vert\left(
\begin{matrix}
\chi_{xx}\mbox{cos}(\theta)+\chi_{xy}\mbox{sin}(\theta)\\
\chi_{yx}\mbox{cos}(\theta)+\chi_{yy}\mbox{sin}(\theta)\\
\end{matrix}
\right)\right\vert,
\end{equation}
where $\vec{M}(h,\,\theta)$ denotes the magnetisation vector in polar coordinates.\\
Results obtained for the angularly averaged susceptibility as a function of $J_y$ and for $J_x=0.9$, $J_z=\gamma=1$ are displayed in Fig. \ref{fig:chi_average} (all the energy scales are expressed in units of $\gamma$). We studied the behavior of such quantity at increasing cluster volume, up to $V=12$ (see the legend in Fig. \ref{fig:chi_average}). Our results are in good agreement with those previously shown in Ref. \cite{rota}, proving the validity of the theoretical framework discussed in the previous sections (please, compare our Fig. \ref{fig:chi_average}, with Fig. 1 and Fig. 4 of Ref.\cite{rota}).\\
Notice that our results are \emph{exact} \ie they have been derived determining the \emph{full} structure of the steady-state configuration without performing any approximation. Furthermore, as far as we know, this is the first time that exact results for the dissipative XYZ model for $V=12$ have been shown. This possibility is mainly due to the reduction of complexity we achieved by means of the algorithm previously described. Indeed, by projecting the dynamics into the relevant submanifold $\mc I _{G}$ we reduced the number of parameters needed by a factor compatible with $\approx 46$ (see Table. \ref{tab:parameter_counting}). Furthermore, notice that when the intensity of the magnetic field is zero, by means of our approach the steady-state configuration for the 12 sites clustes can be found \emph{exactly} by studying a linear problem of size \emph{two order} of magnitude lesser than $\mbox{dim}[\mathbb{H}]^2$ (please compare the value of $\mbox{dim}[\mc I^{+}_{G}]$ to the square of the Hilbert space dimension reported in Table. \ref{tab:parameter_counting}). A discussion about the zero magnetic field case where the generator is parity-preserving is given in the following paragraph.\\

\subsection{Effects of magnetic fields on the steady-state}\label{sec:parity_broken}
\begin{figure}[t!]
\centering
\subfigure{\includegraphics[scale=0.58]{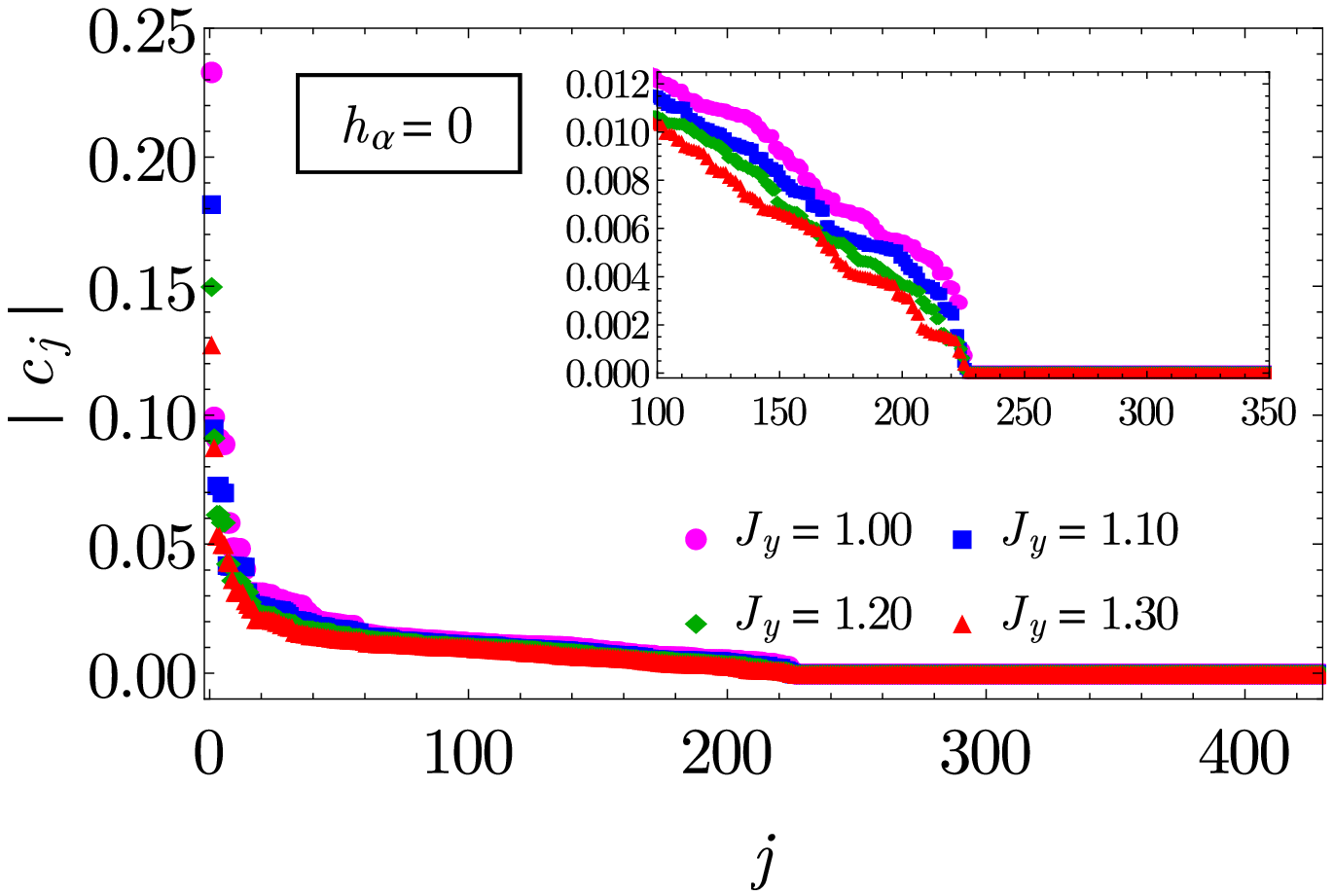}\label{fig:ss_h_zero}}
\subfigure{\hspace{0.3cm}\includegraphics[scale=0.555]{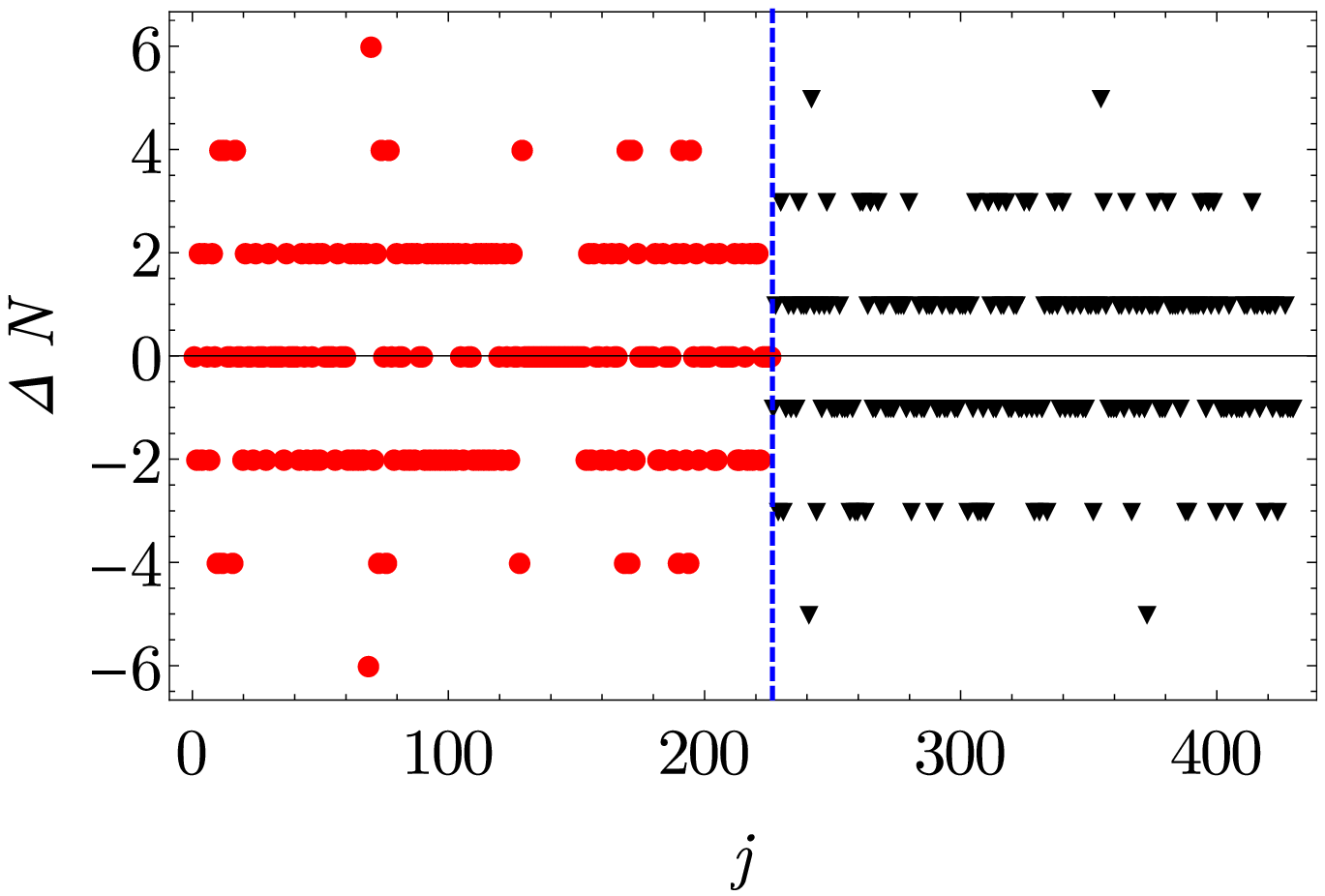}\label{fig:deltan_h_zero}}
\caption{Results for the steady-state configurations of the $2 \times 3$ cluster for $h_x=h_y=0$. Fig. \ref{fig:ss_h_zero}: behavior of the absolute value of the coefficients $c_j$ ($j=1,\cdots,430$) of the steady-state configurations for different values of $J_y$ (see the legend). The inset shows the behavior of the coefficients around $j=226$. Fig \ref{fig:deltan_h_zero}: behavior of $\Delta N$ for the steady-state configuration $J_y=1$ (the other parameters are the same as Fig. \ref{fig:chi_average}). Red circles are associated to states belonging to $\mc I^+_{G}$. Black triangles are associated to odd parity states. The blue dashed vertical line has been added to help locate the boundary between even and odd parity states regions.}\label{fig:rho_zero_mag}
\end{figure}

Let us now consider the structure of the steady-state configuration of the dissipative XYZ model. We first consider the case of zero magnetic field. In such case, the generator $\mc L$ is parity-preserving. Indeed, once the XYZ model is expressed in terms of spin-1/2 ladder operators, it is easy to see that
\be
\mc M^{+}[\rho]= \mc L_{0}[\rho], \quad \mc M^{-} [\rho]=0.
\ee
As a consequence, we expect the steady-state configuration to belong to $\mc I^+ _{G}$. In order to show that this is the case, we first compute the steady-state density matrix $\rho_{ss}$, whose form is given by
\be\label{eq:ss_decomp}
\rho_{ss}= \sum_{j=1}^{\mbox{dim}[\mc I_{G}]} c_j \rho_j,
\ee
and then we consider the behavior of the absolute value of the coefficients $\{c_j\}$. Some results for the $2 \times 3$ cluster ($\gamma=J_z=1$, $J_x=0.9$, $h_x=h_y=0$) are shown in Fig. \ref{fig:rho_zero_mag}, where the absolute values of the coefficients $\{c_j\}$ have been ordered in decreasing order.\\
\begin{figure}[t!]
\centering
\subfigure{\includegraphics[scale=0.58]{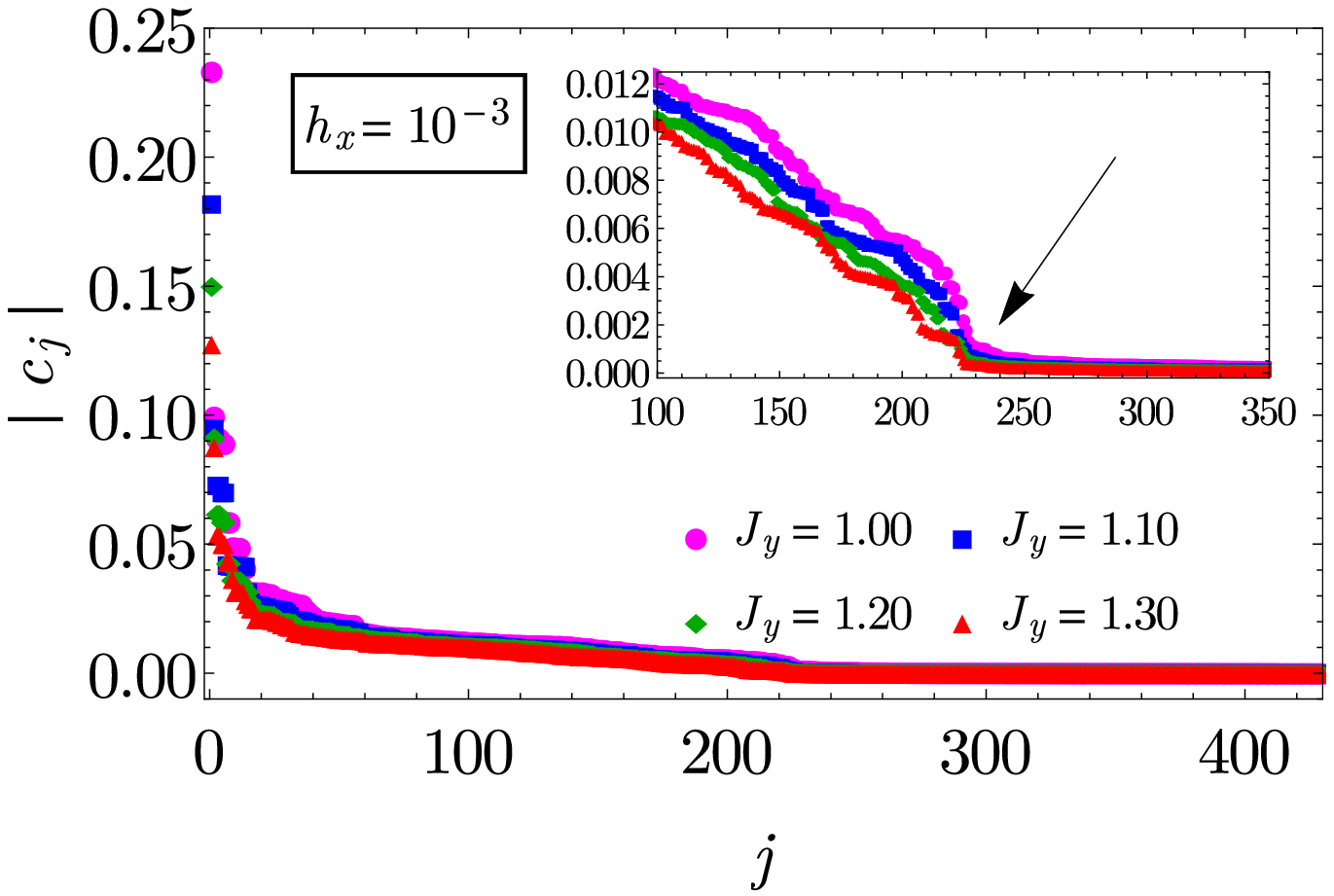}\label{fig:ss_h_nonzero}}
\subfigure{\hspace{0.3cm}\includegraphics[scale=0.555]{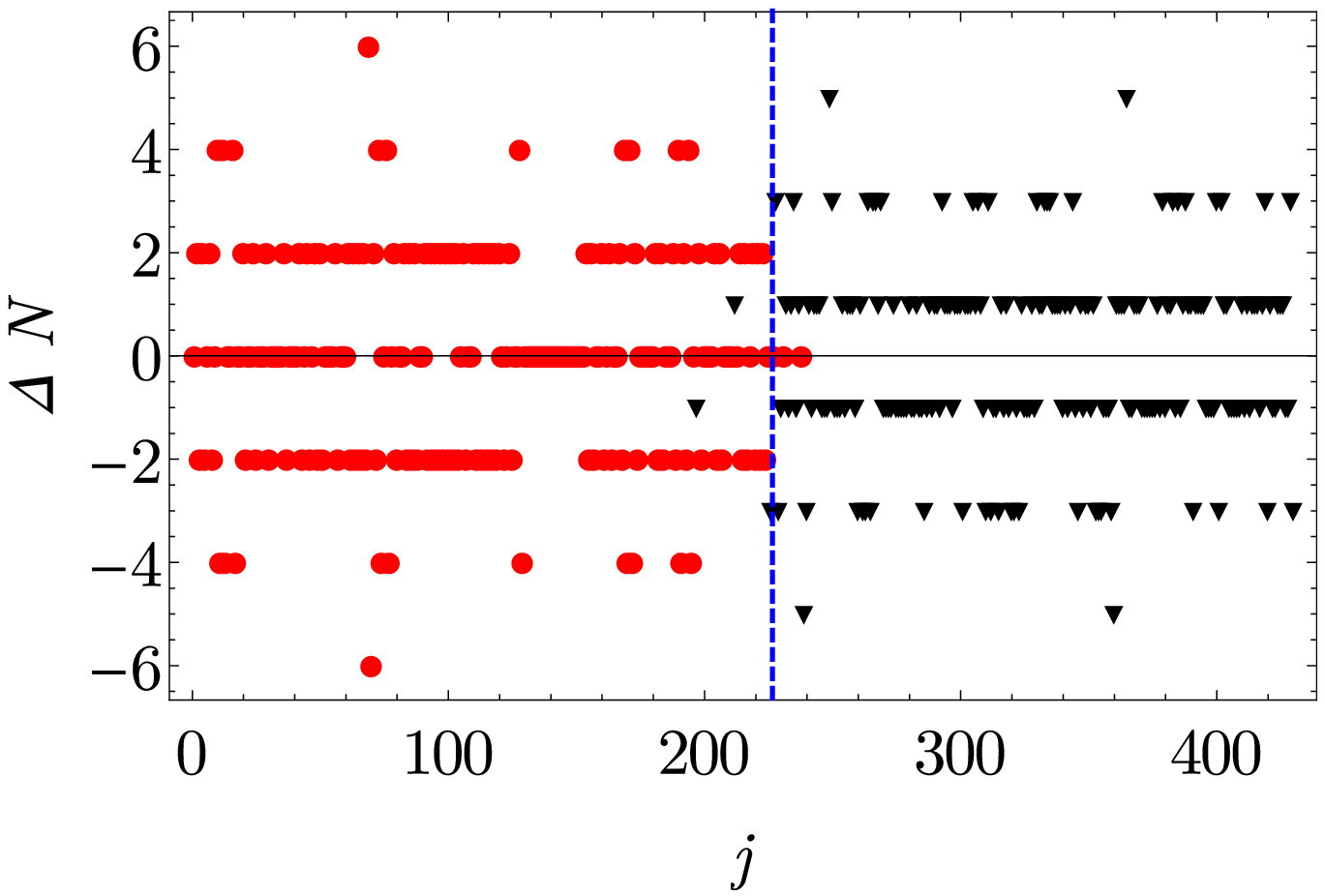}\label{fig:deltan_h_nonzero}}
\caption{Results for the steady-state configurations of the $2 \times 3$ cluster for $h_x=10^{-3}$ and $h_y=0$. Fig. \ref{fig:ss_h_nonzero}: behavior of the absolute value of the coefficients $c_j$ ($j=1,\cdots,430$) of the steady-state configurations for different values of $J_y$ (see the legend). The inset shows the behavior of the coefficients around $j=226$. Fig. \ref{fig:deltan_h_nonzero}: behavior of $\Delta N$ for the steady-state configuration $J_y=1$ (the other parameters are the same as Fig. \ref{fig:chi_average}). Red circles are associated to states belonging to $\mc I^+_{G}$. Black triangles are associated to odd parity states.}\label{fig:rho_non_zero_mag} 
\end{figure}
As it is possible to see in Fig. \ref{fig:ss_h_zero}, for all the values considered only the first 226 coefficients of the steady-state configuration are different from zero. Such number corresponds to the dimension of $\mc I^+_{G}$. This fact can be checked by considering the behavior of $\Delta N$. The results for such quantity for $J_y=1$ are shown in Fig. \ref{fig:deltan_h_zero}. As displayed in Fig.\ref{fig:deltan_h_zero} the first 226 states are those with $\Delta N$ even (red circles), that is those states spannig the subset $\mc I^{+}_G$. Please, notice the separation between opposite parity sectors clearly visible in Fig. \ref{fig:deltan_h_zero}. Similar results have been observed for other lattice clusters and other values of the parameters (not shown).\\

As the magnetic field is switched on, the time-evolution is no more parity preserving. Indeed, in such case a parity mixing term appears in the master equation. In other words, we have that
\be
\mc M^{+}[\rho]=\mc L_0[\rho], \quad \mc M^{-} [\rho]=-i\left[\mc H _m,\,\rho\right].
\ee
As a consequence, we expect the steady-state configuration to acquire tails that go beyond $j=226$ entering in the odd-parity region. Some results for $h_x=10^{-3}$ and $h_y=0$ are shown in Fig. \ref{fig:rho_non_zero_mag}. 
Since the intensity of the magnetic field is still small the overall behavior of the steady-state configuration is quite the same as that previously shown. However, as clearly visible in the inset of Fig. \ref{fig:ss_h_nonzero} (see the arrow), now the coefficients $c_j$ for $j>226$ are non-zero. In addition, as shown in Fig. \ref{fig:deltan_h_nonzero}, altough a separation between opposite parity sectors is still visible, it is no more perfect as in the zero magnetic field case (notice that some odd parity states moved towards lower values of $j$, and some of the even parity states moved toward larger values of $j$).\\

\section{Summary and Conclusions}\label{sec:summary_and_conclusions}
We considered a $d$-dimensional open quantum lattice described by a Lindblad master equation supporting a unique steady-state. We showed that anytime the dynamics is \emph{weakly} symmetric under the action of a group $G$ which is a finite subgroup of the set of all the permutations of the lattice sites, a quest for the unique steady-state can be carried out within a relevant $G$-invariant subspace $\mc I_G$. We provided an explicit algorithm for determining an orthonormal set of such relavant subspace. We explicitly showed that for a spin-1/2 system such subspace has dimension considerably smaller than $\mbox{dim}[\mathbb{H}]^2$. Furthermore, we showed that whenever the evolution is parity preserving, the dynamics is completely determined by the subset $\mc I^+_{G}$ containing the even parity states. Our theoretical findings have been validated by means of numerical simulations for the dissipative XYZ model on two-dimensional clusters. \\
We discussed the validity of our approach by considering the susceptibility properties of such model related to the presence of a non-zero magnetic field in the $xy$-plane. Our results for the angularly averaged susceptibility are compatible with those previously appeared in literature, specifically those appeared in Ref.\cite{rota}. In addition, we showed \emph{exact} results for such model up to 12 sites. As far as we know, the results for the 12 sites cluster have been usually obtained by means of stochastic methods or by means of variational techniques. Here, we demonstrate the efficiency of our approach by determinig the cluster properties exactly \ie determining the \emph{full} structure of the steady-state configuration.\\
The role of parity has been investigated by analysing the steady-state structure with and without a uniform magnetic field in the $xy$-plane. Also in this case, numerics supports our theoretical framework, showing that whenever the time-evolution is parity preserving the steady-state configuration belongs to the subspace of $\mc I _G$ corresponding to even parity states.\\
Altough here we considered only spin-1/2 particles, it is worth noting that our approach can be readily extended to systems with arbitrary spin, to fermion and boson systems (the latter with truncated Fock space to ensure the parameter $B$ to be finite), as well as to hybrid setups involving subparts having different nature.\\
At least in terms of the amount of resources needed, our approach is considerably more efficient then those that determine the steady-state configuration by means of the entire set of projectors on number states. In addition, since $\rho_{ss} \in \mc I _G$ and we provided an explicit algorithm for determining its basis, our findings can be also exploited to determine the exact form of elements that have to be used in variational algorithms. The generalisation to the case of a reducible open dynamics \ie in the presence of non-trivial conserved quantities, and results for larger clusters will be shown in future works.\\
\section*{Acknowledgments}
The author is very grateful to Martino De Leo for the multiple and helpful discussions. The author thanks also Davide Rossini for carefully reading the manuscript when it was in preparation.\\

\end{document}